\documentclass[11pt]{article}

\usepackage{bm}
\usepackage{amsmath,amssymb}
\usepackage{graphicx}

\begin{document}
\title{\textbf{Non-Gibrat's law in the middle scale region}}

\author{
\\
Masashi Tomoyose  $^{1}$\footnote{e-mail address: mtomo@weiss.sci.u-ryukyu.ac.jp}, 
Shouji Fujimoto $^{2}$\footnote{e-mail address: fujimoto@kanazawa-gu.ac.jp}
and Atushi Ishikawa $^{2}$\footnote{e-mail address: ishikawa@kanazawa-gu.ac.jp}
\\ \\
$^1$University of the Ryukyus, 903-0213, Japan\\
$^2$Kanazawa Gakuin University, 920-1392, Japan\\
}
\date{}
\maketitle

\begin{abstract}
By using numerical simulation,
we confirm that Takayasu--Sato--Takayasu (TST) model 
which leads Pareto's law satisfies the detailed balance under Gibrat's law.
In the simulation, we take an exponential tent-shaped function as the growth rate distribution.
We also numerically confirm the reflection law
equivalent to
the equation which gives the Pareto index $\mu$ in TST model.
Moreover, we extend the model modifying the stochastic coefficient 
under a Non-Gibrat's law.
In this model, the detailed balance is also numerically observed.
The resultant pdf is power-law in the large scale Gibrat's law region,
and is the log-normal distribution in the middle scale Non-Gibrat's one.
These are accurately confirmed in the numerical simulation.
\end{abstract}

\begin{flushleft}
 PACS code : 89.65.Gh
\end{flushleft}

\section{Introduction}

Recent developments of computer technology enable vast quantities of economic data
to be analyzed.
As one of remarkable examples,
Fujiwara et al. \cite{FSAKA} find that
Pareto's law \cite{Pareto} 
\begin{eqnarray}
    P(x) \propto x^{-(\mu+1)}~~~~{\rm for }~~~~x > x_0
    \label{Pareto}
\end{eqnarray}
can be derived 
from the law of detailed balance
\begin{eqnarray}
    P_{1 2}(x_1, x_2) = P_{1 2}(x_2, x_1)~
    \label{Detailed balance}
\end{eqnarray}
and Gibrat's law \cite{Gibrat}
\begin{eqnarray}
    Q(R|x_1) = Q(R)~,
    \label{Gibrat}
\end{eqnarray}
which are observed in massive amounts of economic data.
Here $x$ is wealth, income, 
profits, assets, sales, the number of employees and so forth,
and $x_1$, $x_2$ are two successive those.
$P_{1 2}(x_1, x_2)$ is a joint probability density function (pdf), and
$Q(R|x_1)$ is a conditional pdf of the growth rate
$R = x_2/x_1$~.

The Pareto's law (\ref{Pareto}) is not observed below some threshold $x_0$~\cite{Gibrat, Badger},
because the Gibrat's law (\ref{Gibrat}) does not hold 
in the middle scale region \cite{Stanley1, TTOMS, Aoyama}. 
In Ref.~\cite{Ishikawa3},
we show that the detailed balance (\ref{Detailed balance}) 
and a Non-Gibrat's law
\begin{eqnarray}
    Q(R|x_1)&=&Const.~R^{\mp t_{\pm}(x_1)-1}~~~~~{\rm for}~~R \gtrless 1~
    \label{tent-shaped1}\\
    t_{\pm}(x_1)&=&t_{\pm}(x_0) \pm \alpha~\ln \frac{x_1}{x_0}~
    \label{Non--Gibrat}
\end{eqnarray}
lead the log-normal distribution in the middle scale region as follows:
\begin{eqnarray}
    P(x_1) = C {x_1}^{-\left(\mu+1\right)}~e^{-\alpha \ln^2 \frac{x_1}{x_0}}~
    ~~~~{\rm for }~~~~x_{\rm min} < x_1 < x_0.
    \label{HandM}
\end{eqnarray}
The Non-Gibrat's law means a statistical dependence of the growth rate distribution 
on the past value $x_1$
in the middle scale region. 
In the derivation, 
we employ the approximation (\ref{tent-shaped1})
observed in profits data of Japanese firms\cite{TSR}, and
find that the expression of $t_{\pm}(x_1)$ (\ref{Non--Gibrat}) is unique
under the detailed balance.
Employing such empirical data, we confirm 
that
the resultant distribution (\ref{HandM}) fits with data in the large scale Pareto's law region
and those in the middle scale log-normal region consistently.
The parameters are estimated as follows: 
$\alpha \sim 0$ for $x_1 > x_0$, 
$\alpha \sim 0.14$ for $x_{\rm min} < x_1 < x_0$,\footnote{
Here, a constant parameter $\alpha$ takes different values in two regions.
This is not an exact procedure.
However, for firms which are in the large scale region in both years ($x_1 > x_0$ and $x_2 > x_0$) 
or in the middle scale one ($x_{\rm min} < x_1 < x_0$ and $x_{\rm min} < x_2 < x_0$),
this procedure is exact. In the database,
most firms stay in the same region.
This parameterization is, therefore, approximately valid for
describing the probability density function $P(x_1)$.
This is empirically confirmed in Ref.~\cite{Ishikawa3}.
}
$x_0 \sim 63,000$ thousand yen and
$x_{\rm min} \sim 1,600$ thousand yen \cite{Ishikawa3}.

These data analyses are sufficiently rigorous,
but at the same time are restricted in the finite period and category,
because it is not easy to procure exhaustive data.
If we build a model based on the detailed balance and (Non-)Gibrat's law,
various economic situations can be simulated.
This might make the reason clear why the parameters take the empirical values. 
Furthermore, we can study assets or sales data of firms which are difficult to obtain.
In this paper, we will propose the simulation model 
based on the observation of economic data.

\section{TST model}

Firstly, we identify the model which 
leads Pareto's law
based on the detailed balance and Gibrat's law in the large scale region.
One of the simplest and powerful candidates is Takayasu--Sato--Takayasu (TST) model \cite{TST},
which is given by the Langevin equation
\begin{eqnarray}
    x(t+1) = b(t) x(t) + f(t)~,
    \label{Langevin equation}        
\end{eqnarray}
where $b(t)$ is a non-negative stochastic coefficient and $f(t)$ is a random
additive noise. 
They show that the conditions
\begin{eqnarray}
    \langle \ln b(t) \rangle < 0, \langle b(t)^2 \rangle > 1
    \label{conditions}
\end{eqnarray}
are necessary and sufficient for the power-law (\ref{Pareto}),
the index $\mu$ of which is given by the equation 
\begin{eqnarray}
    \langle b(t)^{\mu} \rangle = 1~.
    \label{equation}
\end{eqnarray}
Here $\langle \cdots \rangle$ denotes an average over realizations.

\begin{figure}[htb]
 \begin{minipage}[t]{0.47\textwidth}
  \includegraphics[width=7.1 cm]{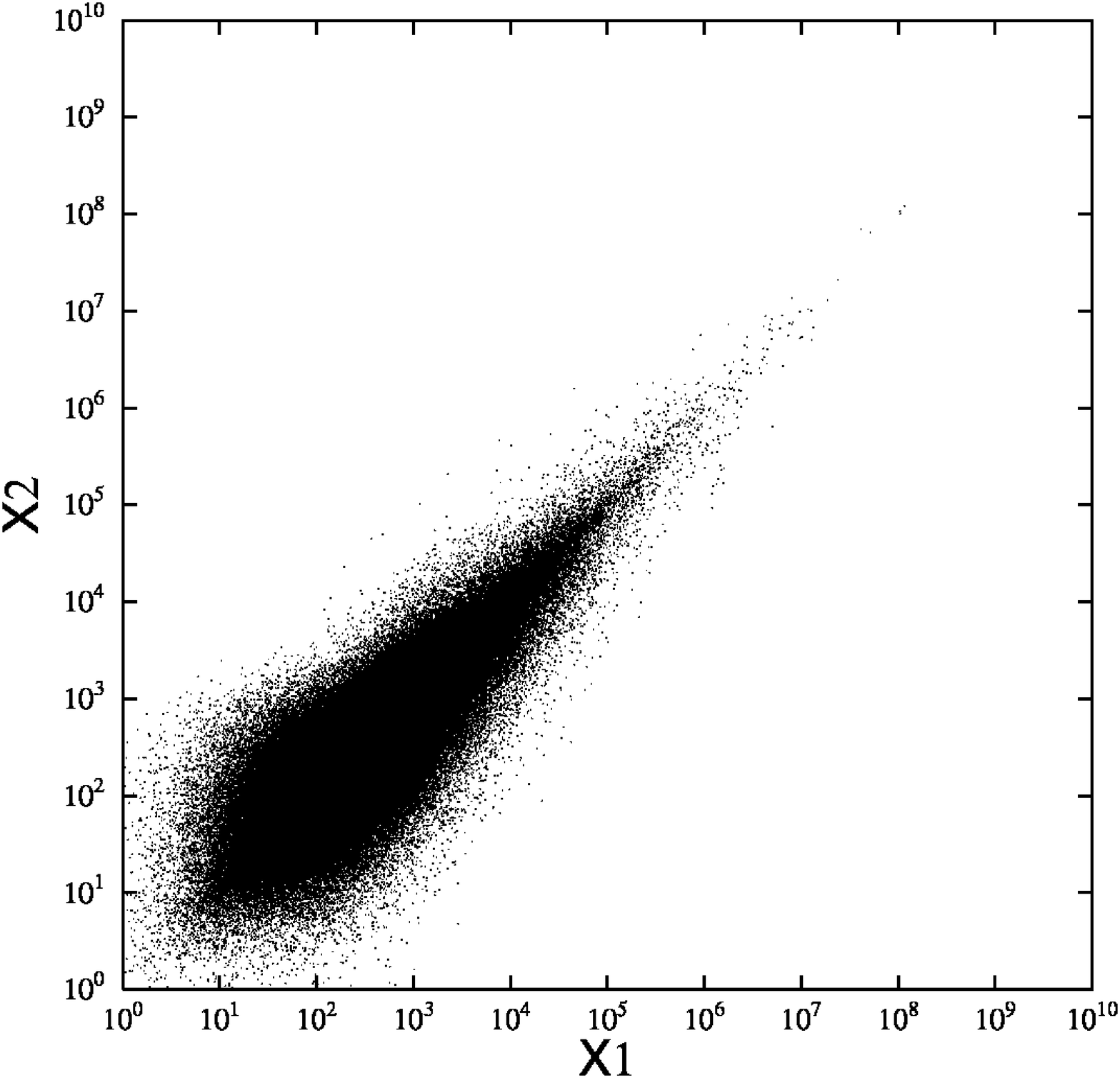}
 \caption{The scatter plot of data points under Gibrat's law,
 the number of which is ``500,000 ".}
 \label{SimulationDB}
 \end{minipage}
 \hfill
 \begin{minipage}[t]{0.47\textwidth}
  \includegraphics[width=7.1 cm]{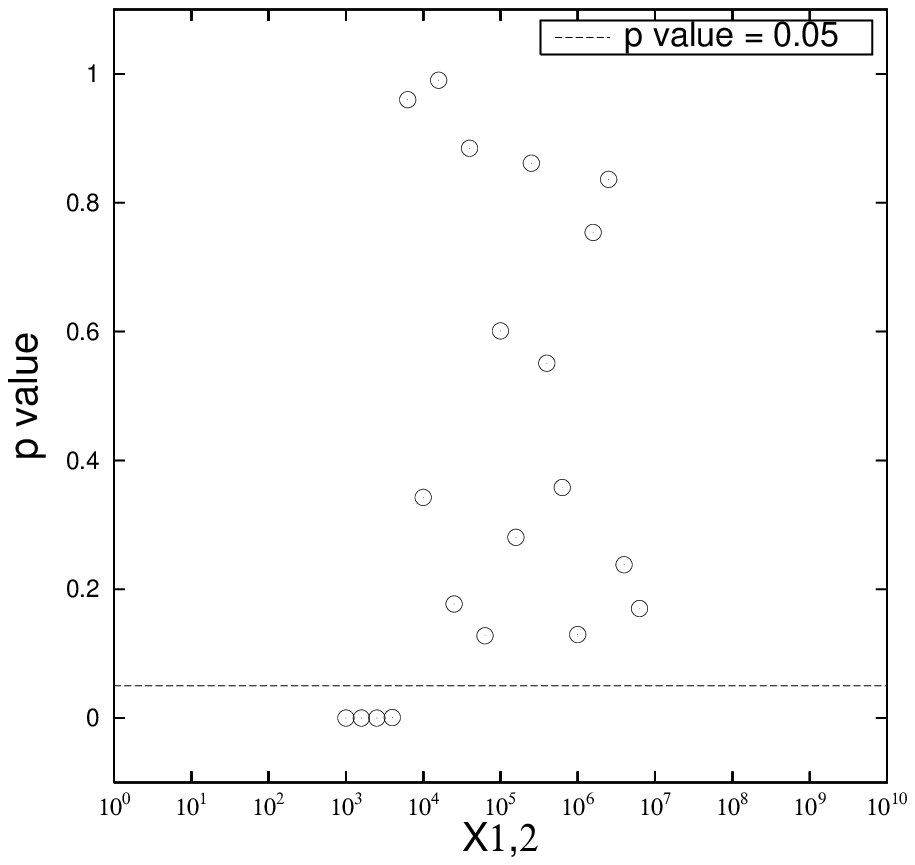}
 \caption{Each $p$ value of the one-dimensional Kolmogorov-Smirnov test 
 for Fig.~\ref{SimulationDB}.}
 \label{SimulationKS}
 \end{minipage}
\end{figure}

\begin{figure}[htb]
 \begin{minipage}[tb]{0.47\textwidth}
  \includegraphics[width=7.1cm]{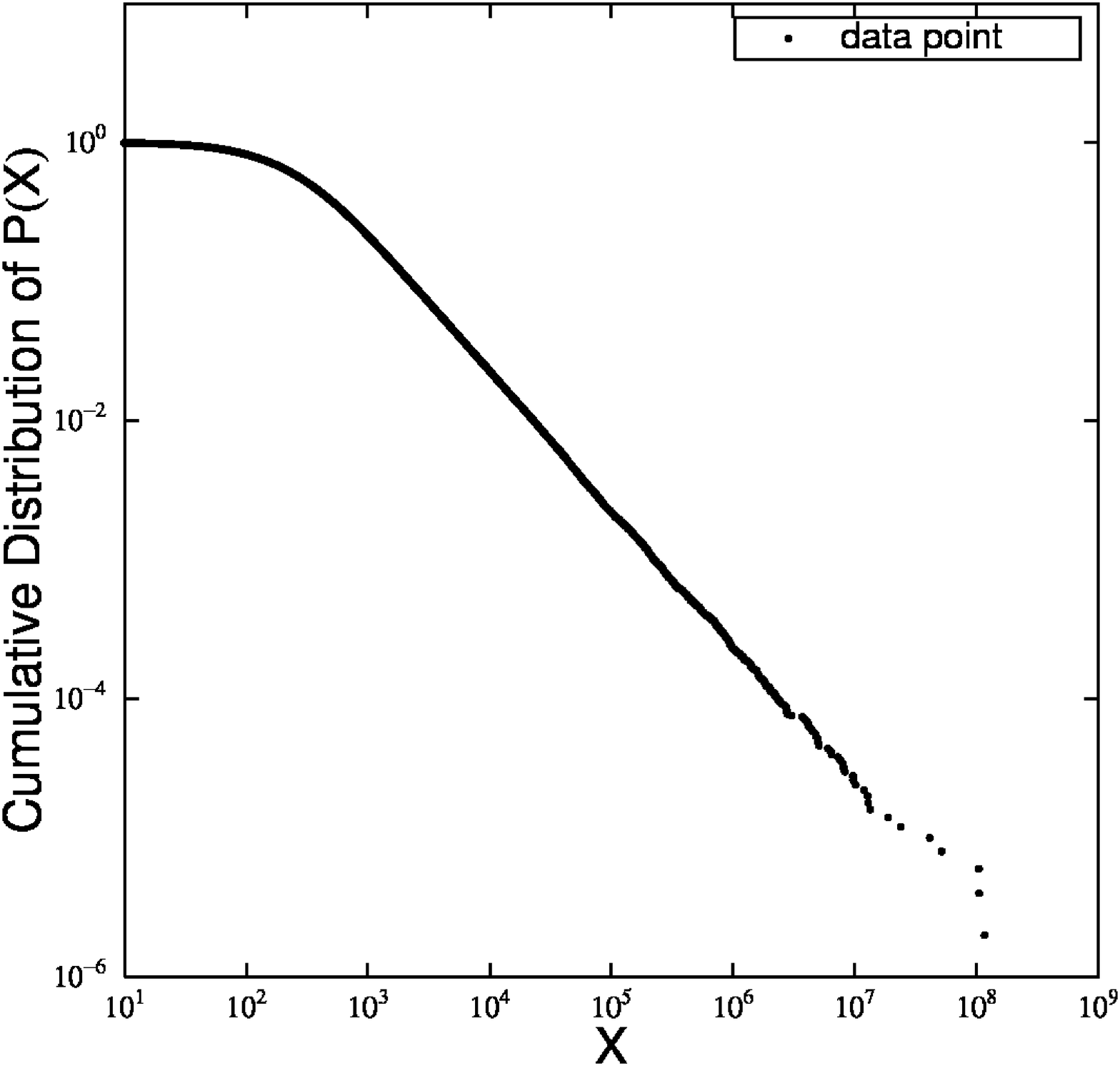}
 \caption{The cumulative distribution of $P(x)$ shows the Pareto's law above 
 $x_{\rm min} \sim 10^4$.
 The Pareto index $\mu$ is estimated by nearly $1$.}
 \label{SimulationCumDist}
 \end{minipage}
\hfill
 \begin{minipage}[tb]{0.47\textwidth}
  \includegraphics[width=7.1cm]{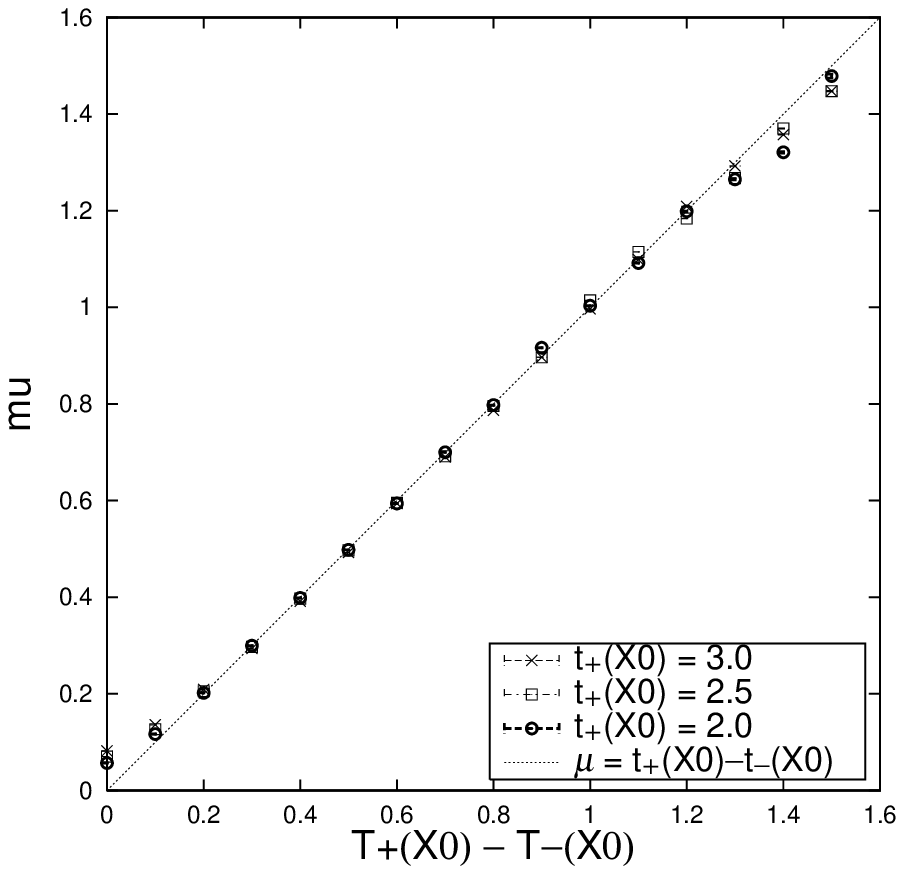}
 \caption{The relation between $\mu$ and $t_{+}(x_0) - t_{-}(x_0)$.
 Here, $\mu$ is estimated by the least square method of the simulation result,
 in which we take various values of $t_{\pm}(x_0)$.}
 \label{tpx0-tmx0-mu}
 \end{minipage}
\end{figure}

In order to claim that this model is consistent with the empirical observation,
the detailed balance and Gibrat's law must be satisfied. From
Eq.~(\ref{Langevin equation}), 
Gibrat's law is hold in the region where a noise $f(t)$ is negligible.
The result in Ref.~\cite{FSAKA} suggests, therefore, that there is the
detailed balance in TST model which leads Pareto's law.
In the numerical simulation, we adopt
the distribution of $b(t)$ and $f(t)$ 
to be Eqs.~(\ref{tent-shaped1})--(\ref{Non--Gibrat}) with $\alpha=0$ 
and Weibull distribution: 
$\frac{k}{\lambda}(\frac{x}{\lambda})^{k-1}
\exp[-(\frac{x}{\lambda})^k]$, respectively. 
The typical scatter plot of the simulation is shown in Fig.~\ref{SimulationDB},
where we take $t_+(x_0)=2.5$, $t_-(x_0)=1.5$, $x_0=10^0$, $k=0.5$ and $\lambda=30$.
To check the validity of the detailed balance, we take
the one-dimensional Kolmogorov-Smirnov (K-S) test.
We compare the distribution sample for 
$P(x_1 \in [10^{3+0.2 (n-1)}, 10^{3+0.2 n}],x_2)$ 
with another sample for $P(x_1,x_2\in [10^{3+0.2 (n-1)}, 10^{3+0.2 n}])$
with $n=1,2, \cdots, 20$
by making the null hypothesis that these two samples are taken from a
same parent distribution. 
Each $p$ value is shown in Fig.~\ref{SimulationKS}.
The null hypothesis is not rejected in 5$\%$ significance level
in the region where the noise $f(t)$ is negligible.
We recognize that the detailed balance is observed 
in the region.
The lower bound of the region is denoted by $x_{\rm min}$, above which
Pareto's law is observed (Fig.~\ref{SimulationCumDist}).

In addition, we also confirm the reflection law
    $Q(R) = R^{-\mu-1} Q(R^{-1})$
derived in Ref.~\cite{FSAKA}.
In this simulation, 
from Eqs.(\ref{tent-shaped1}) and (\ref{Non--Gibrat}) with $\alpha=0$, 
the reflection law is reduced to \footnote{
This equation is also confirmed empirically
in profits data of Japanese firms \cite{Ishikawa3}.}
\begin{eqnarray}
    \mu = t_{+}(x_0) - t_{-}(x_0),
    \label{reflection law}
\end{eqnarray}
which is equivalent to the condition (\ref{equation}). 
We observe this reflection law in various $t_{\pm}(x_0)$ (Fig.~\ref{tpx0-tmx0-mu}). From these results,
TST model is appropriate satisfying the detailed balance under the Gibrat's law.

\section{Simulation under Non-Gibrat's law}
If 
the detailed balance is satisfied even under a Non-Gibrat's law in the simulation,
the log-normal distribution must be reduced in the middle scale region
\cite{Ishikawa3}. 
In a stochastic method,
this scheme is not obvious analytically.
Because a non-negative stochastic coefficient in Eq.~(\ref{Langevin equation})
must be modified as $b(x(t), t)$
under the Non-Gibrat's law.
This extension exceeds the analytical framework of TST model.
However, we can examine the scheme numerically.

\begin{figure}[htb]
 \begin{minipage}[htb]{0.45\textwidth}
 \centerline{\includegraphics[width=6 cm, height=5.3 cm]{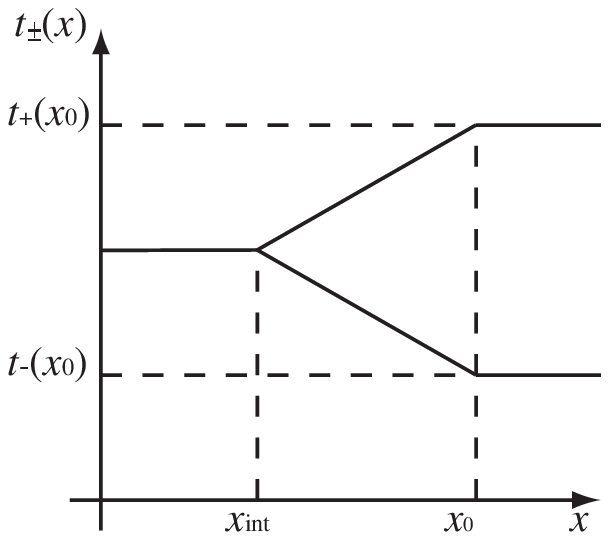}}
 \caption{Continuous functions 
$t_{\pm}(x)= t_{\pm}(x_0) \pm \alpha \ln \frac{x}{x_0}$
 with $\alpha = 0$ for $x > x_0$, $\alpha > 0$ for $x_{\rm int} < x < x_0$
 and $\alpha = 0$ for $x < x_{\rm int}$.
 The horizontal axis is logarithmic scale.}
 \label{setumei01}
 \end{minipage}
 \hfill
 \begin{minipage}[htb]{0.45\textwidth}
 \centerline{\includegraphics[width=7.8 cm, height=5.1
  cm]{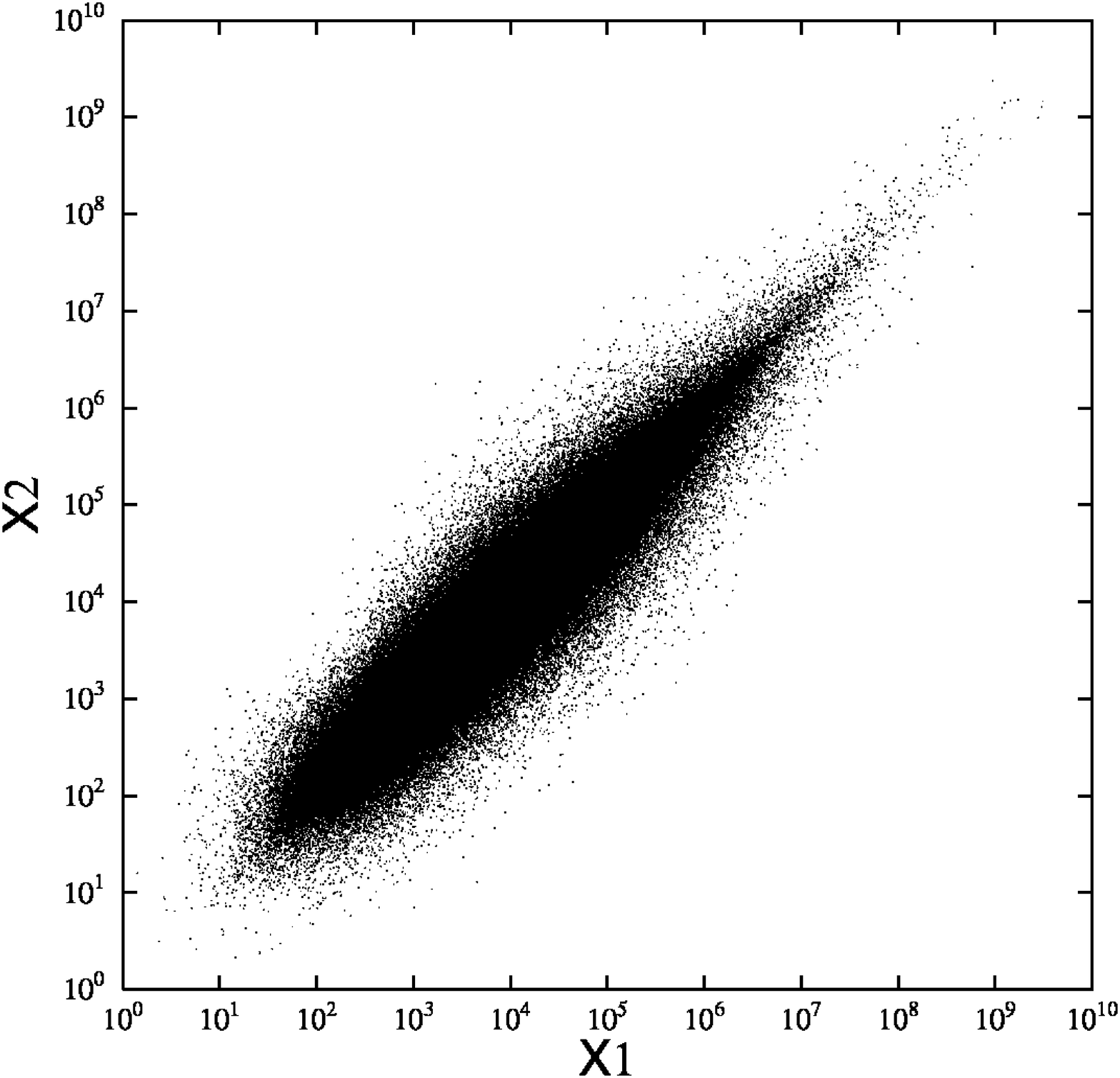}}
 \caption{The scatter plot of data points under Non-Gibrat's law,
 the number of which is ``500,000 ".}
 \label{SimulationDB2}
 \end{minipage}
\end{figure}

In the simulation,
$x_{\rm int}$ is introduced satisfying $t_{+}(x_{\rm int}) = t_{-}(x_{\rm int})$.
We adopt the distribution of $b$ to be Eqs.~(\ref{tent-shaped1})--(\ref{Non--Gibrat})
with $\alpha = 0$ for $x > x_0$ (Gibrat's law), $\alpha \neq 0$ for $x_{\rm int} < x < x_0$
(Non-Gibrat's law)
and $\alpha = 0$ for $x < x_{\rm int}$
keeping $t_{\pm}(x)$ continuously (Fig.~\ref{setumei01}).
The last parameterization is imposed to exclude immoderate slopes of growth rate distributions.
As the Non-Gibrat's region in the middle scale, we observe the region 
$\max(x_{\rm min}, x_{\rm int}) < x < x_0$.
\begin{figure}[htb]
 \begin{minipage}[t]{0.41\textwidth}
  \includegraphics[width=7.8 cm, height=5.3 cm]{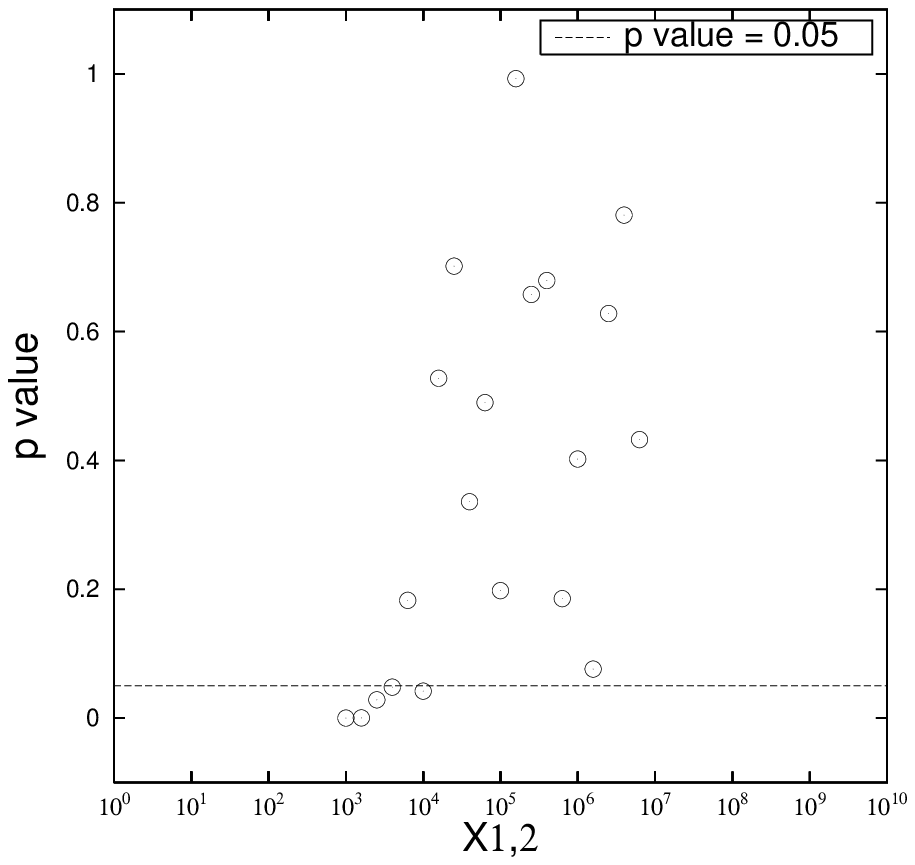}
 \caption{The result of one-dimensional Kolmogorov-Smirnov test for Fig.~\ref{SimulationDB2}.}
 \label{SimulationKS2}
 \end{minipage}
 \hfill
 \begin{minipage}[t]{0.54\textwidth}
\hspace*{8mm}
  \includegraphics[width=7.8 cm, height=5.3 cm]{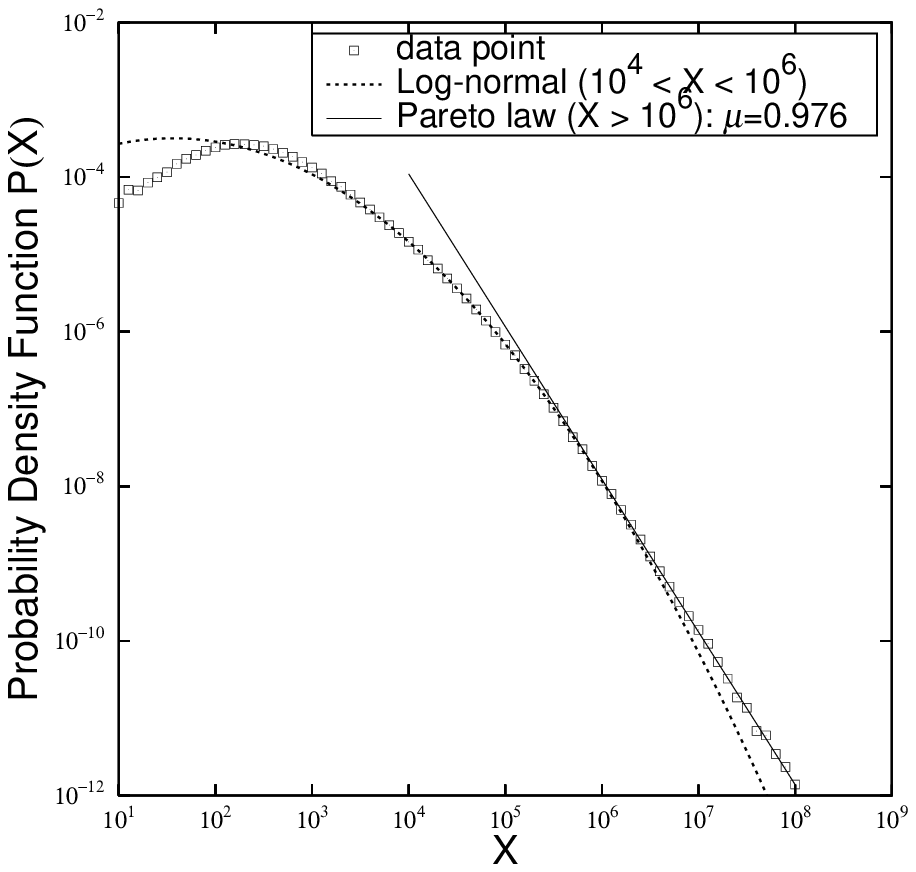}
 \caption{The resultant pdf $P(x)$ with $t_+(x_0)=2.5$, $t_-(x_0)=1.5$, $x_0=10^6$, $k=0.5$,
 $\lambda=30$
 and $\alpha=0.1$. In the large scale region,
 Pareto's law is observed. In the middle scale one, the log-normal distribution
 is observed.
 }
 \label{SimulationDist}
 \end{minipage}
\end{figure}

\begin{figure}[htb]
 \begin{minipage}[htb]{0.45\textwidth}
  \includegraphics[width=7.8 cm, height=5.1 cm]{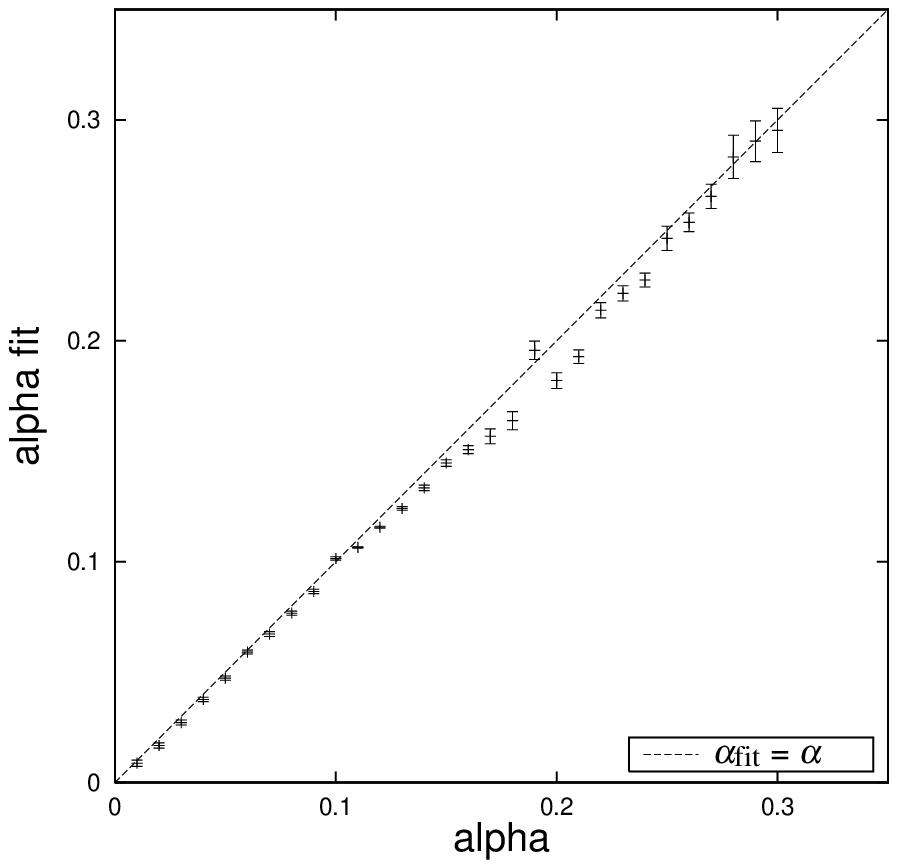}
 \caption{The relation between $\alpha$ and $\alpha_{\rm fit}$
 for the case $t_+(x_0)=2.5$, $t_-(x_0)=1.5$, $x_0=10^6$, $k=0.5$ and $\lambda=30$.}
 \label{alpha-alpha}
 \end{minipage}
 \hfill
 \begin{minipage}[htb]{0.45\textwidth}
  \includegraphics[width=7.8 cm, height=5.1 cm]{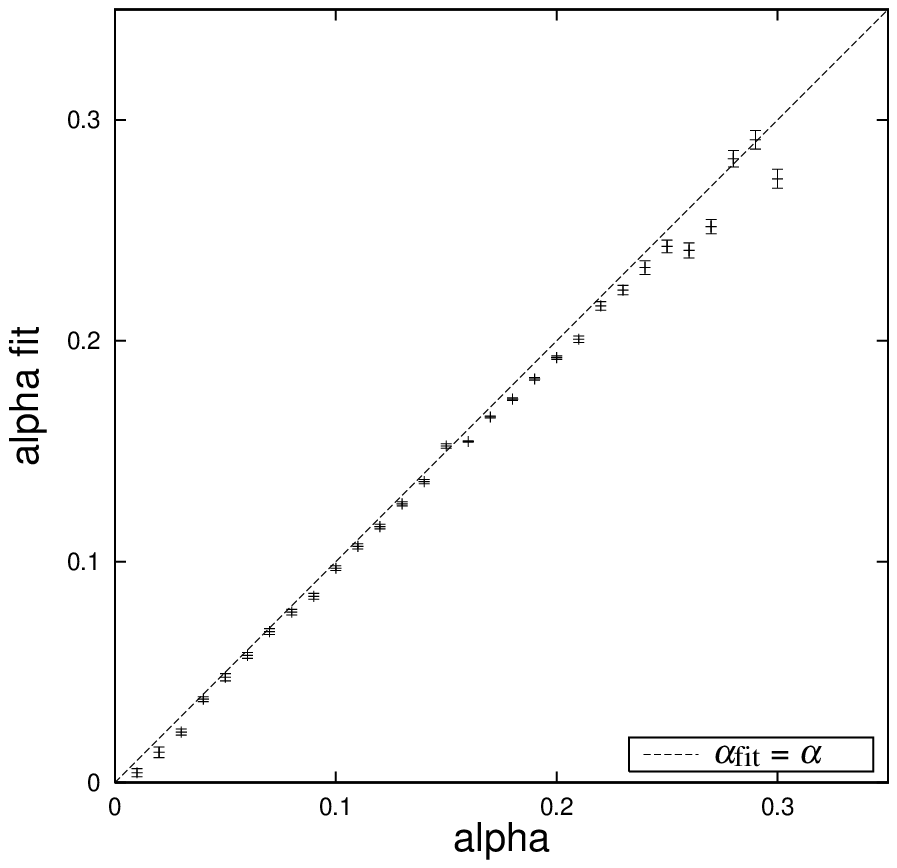}
 \caption{The relation between $\alpha$ and $\alpha_{\rm fit}$
 for the case $t_+(x_0)=3.5$, $t_-(x_0)=2.0$, $x_0=10^6$, $k=0.5$ and $\lambda=30$.}
 \label{alpha-alpha2}
 \end{minipage}
\end{figure}

The typical scatter plot of the simulation is shown in Fig.~\ref{SimulationDB2},
where we take $t_+(x_0)=2.5$, $t_-(x_0)=1.5$, $x_0=10^6$, $k=0.5$, $\lambda=30$ 
and $\alpha=0.1$.
To check the validity of the detailed balance, we take
the same K-S test in the previous section.
The result is shown in Fig.~\ref{SimulationKS2}.
The detailed balance is confirmed not only in the Gibrat's large scale region
($x>x_0=10^6$) but also in the Non-Gibrat's middle scale one
($\max(x_{\rm min}, x_{\rm int})=10^4 < x < x_0=10^6$). 
The resultant pdf of $x$ is shown in Fig.~\ref{SimulationDist},
where the log-normal distribution in the middle scale region is observed
in addition to Pareto's law in the large scale one.
To confirm the validity of the log-normal distribution,
we check the correlation between the parameter $\alpha$ inputted in the simulation and that estimated by fitting 
the pdf in the Non-Gibrat's region (denoted by $\alpha_{\rm fit}$).
The result is shown in Fig.~\ref{alpha-alpha} for the case 
$t_+(x_0)=2.5$, $t_-(x_0)=1.5$, $x_0=10^6$, $k=0.5$ and $\lambda=30$ and
in Fig.~\ref{alpha-alpha2} for the case
$t_+(x_0)=3.5$, $t_-(x_0)=2.0$, $x_0=10^6$, $k=0.5$ and $\lambda=30$ for example.
The correlation is very high.
Consequently, this simulation model is also
appropriate satisfying the detailed balance even under the Non-Gibrat's
law.
\section{Summary and future problems}

In this study, by using numerical simulation,
we confirm that TST model which leads Pareto's law satisfies the detailed balance under Gibrat's law.
In the simulation, we take an exponential tent-shaped function as the growth rate distribution.
We also numerically confirm the reflection law \cite{FSAKA}
equivalent to
the equation which gives the Pareto index $\mu$ in TST model.
Moreover, we extend the model modifying the stochastic coefficient under a Non-Gibrat's law.
In this model, the detailed balance is also observed.
The resultant pdf is power-law in the large scale Gibrat's law region,
and is the log-normal distribution in the middle scale Non-Gibrat's one.
These are accurately confirmed in the numerical simulation.

\begin{figure}[htb]
 \begin{minipage}[htb]{0.45\textwidth}
 \centerline{\includegraphics[width=45 mm]{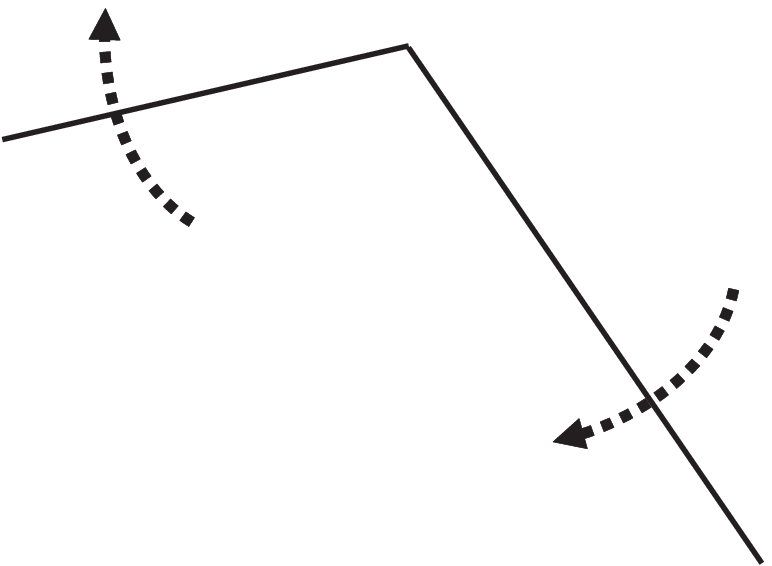}}
 \caption{In the middle scale region, 
 the probability of the positive growth decreases and 
 the probability of the negative growth increases
 symmetrically as the classification of $x$ increases.}
 \label{test}
 \end{minipage}
 \hfill
 \begin{minipage}[htb]{0.45\textwidth}
 \centerline{\includegraphics[width=45 mm]{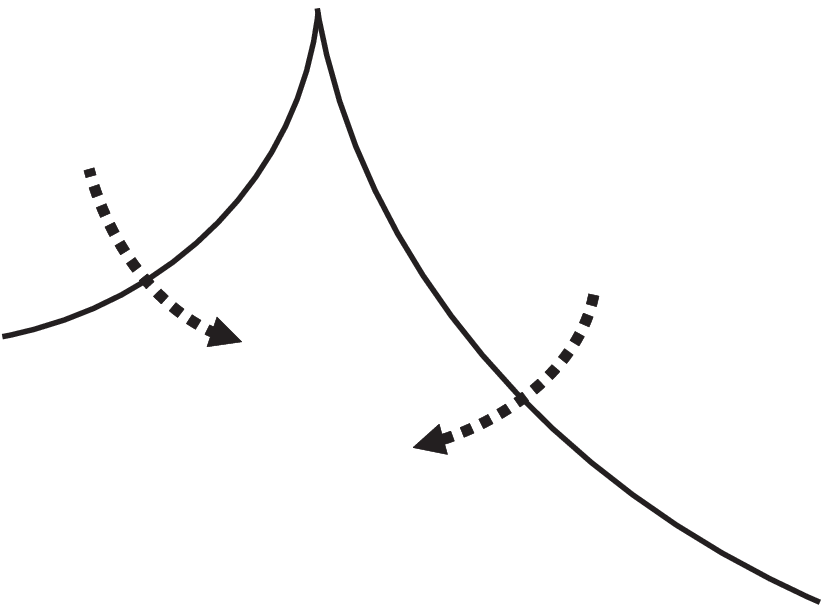}}
 \caption{As for sales or assets of firms, 
 the probability of the positive and negative growth
 decrease simultaneously as the classification of $x$ increases.}
 \label{test2}
 \end{minipage}
\end{figure}

In this simulation, we employ the Non-Gibrat's law (\ref{Non--Gibrat}) that
the probability of the positive growth decreases and that of the negative growth increases
symmetrically as the classification of $x$ increases 
(Figs.~\ref{setumei01},~\ref{test}).
This is uniquely derived from the exponential tent-shaped form (\ref{tent-shaped1})
observed in Japanese profits data \cite{Ishikawa3}.
On the other hand, 
as for sales or assets of firms,  
it is reported that
the probability of the positive and negative growth
decrease simultaneously
as the classification of $x$ increases (Fig.~\ref{test2}). 
This difference is thought to be caused by the form of the exponential tent-shaped 
growth rate distribution.
Eq.~(\ref{tent-shaped1}) is rewritten 
by using the pdf $q(r|x_1)$ for $r=\log_{10} R$ as follows:
\begin{eqnarray}
    \log_{10}q(r|x_1) = c \mp t_{\pm}(x_1)~r~~~~~{\rm for}~~r \gtrless 0~.
\end{eqnarray}   
For sales or assets of firms,  
these linear approximations are not used to express the growth rate
distribution with the curvature \cite{FSAKA}.

In order to count the curvature,
we add a second order term with respect to $r$:
\begin{eqnarray}
    \log_{10}q(r|x_1)&=&c \mp t_{\pm}(x_1)~r+\ln10~u_{\pm} (x_1)~r^2
    ~~~~~{\rm for}~~r \gtrless 0~.
    \label{approximation1}
\end{eqnarray}
In terms of $Q(R|x_1)$, these are expressed as
\begin{eqnarray}
    Q(R|x_1)=d~R^{-1 \mp t_{\pm}(x_1)+u_{\pm}(x_1) \ln R}~~~~~{\rm for}~~R \gtrless 1~.
    \label{non-tent-shaped1}
\end{eqnarray}
In this case,
the detailed balance (\ref{Detailed balance}) is reduced to be
\begin{eqnarray}
    \frac{P(x_1)}{P(x_2)} = \frac{1}{R} \frac{Q(R^{-1}|x_2)}{Q(R|x_1)}~
    = R^{~1+t_{+}(x_1) - t_{-}(x_2)-\left[u_+(x_1)-u_-(x_2)\right]\ln R}~
    \label{start}    
\end{eqnarray}
for $R>1$.
By setting $R=1$ after
differentiating Eq.~(\ref{start}) with respect to $R$, 
differential equations are obtained.
The solutions are uniquely fixed as
\begin{eqnarray}
    t_+(x) &=& t_+(x_0) + \alpha \ln \frac{x}{x_0}
                + \frac{\beta}{2} \ln^2 \frac{x}{x_0}
                 + \frac{\gamma}{3} \ln^3 \frac{x}{x_0}~,
    \label{t+}\\
    t_-(x) &=& t_-(x_0) - (\alpha - \eta) \ln \frac{x}{x_0}
                - \frac{\beta - \delta }{2} \ln^2 \frac{x}{x_0}
                  - \frac{\gamma}{3} \ln^3 \frac{x}{x_0}~,
    \label{t-}\\
    u_+(x) &=& u_+(x_0) 
                - \frac{\beta + \delta}{6} \ln \frac{x}{x_0}
                - \frac{\gamma}{6} \ln^2 \frac{x}{x_0}~,
    \label{u+}\\
    u_-(x) &=& u_-(x_0) 
                - \frac{\beta - 2 \delta}{6} \ln \frac{x}{x_0}
                - \frac{\gamma}{6} \ln^2 \frac{x}{x_0}~,
    \label{u-}\\
    P(x) &=& C~x^{-(\mu+1)}~
    \exp \left[{-(\alpha-\frac{\eta}{2})\ln^2 \frac{x}{x_0}
    -\frac{2 \beta - \delta}{6}\ln^3 \frac{x}{x_0}
    -\frac{\gamma}{6}\ln^4 \frac{x}{x_0}} \right]~
    \label{P}    
\end{eqnarray}
with $\eta/2 = -u_+(x_0) + u_-(x_0)$~.
Same solutions are obtained for $R<1$~.
Here we set the parameters as follows:
$\alpha, \beta, \gamma, \delta, \eta = 0$ for $x > x_0$ (Gibrat's law),
$\alpha, \beta, \gamma, \delta \neq 0$ and $\eta=0$ for $x < x_0$ (Non-Gibrat's law)
because $u_{\pm}(x_0)$ are constants.

\begin{figure}[htb]
 \begin{minipage}[htb]{0.45\textwidth}
 \centerline{\includegraphics[width=5.0 cm, height=5.0 cm]{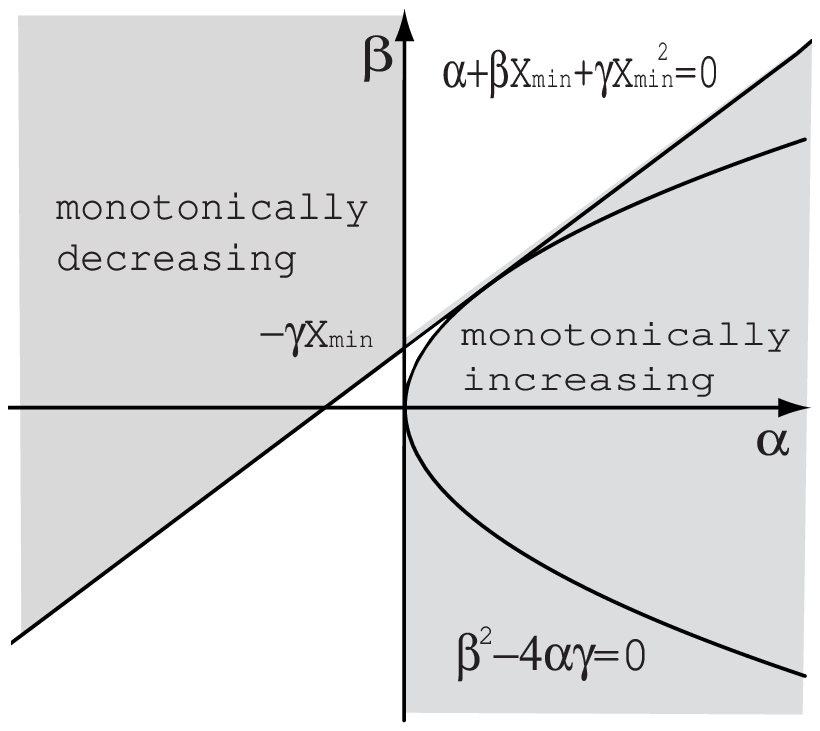}}
 \caption{For $\gamma>0$, the monotonically increasing or decreasing
 regions of $t_+(x)$~.}
 \label{fig05}
 \end{minipage}
 \hfill
 \begin{minipage}[htb]{0.45\textwidth}
 \centerline{\includegraphics[width=5.0 cm, height=5.0 cm]{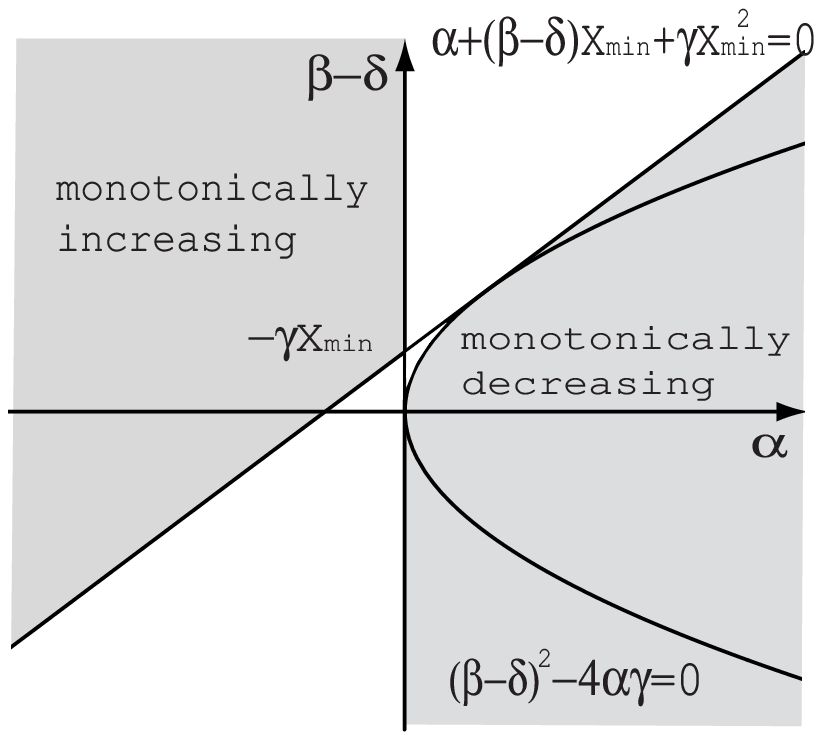}}
 \caption{For $\gamma>0$, the monotonically increasing or decreasing
 regions of $t_-(x)$~.}
 \label{fig06}
 \end{minipage}
\end{figure}

As an index of increase and decrease of the probability of the growth rate,
we examine the differential coefficient at the origin: 
${\frac{d}{d r} \log_{10} q(r|x_1)}_{r \to \pm 0} = \mp t_{\pm} (x_1)$~.
We are, therefore, interested in the increase and decrease of $t_{\pm}(x)$ 
in the middle scale region.
In the case $\gamma>0$ for example,
the results are shown in Figs.~\ref{fig05} and \ref{fig06}
using the notation
$X \equiv \ln (x/x_0)$, $X_{\rm min} \equiv \ln (x_{\rm min}/x_0)$~.
Consequently the solution, that
the probability of the positive and negative growth
decrease simultaneously as the classification of $x$ increases,
exists in following parameter regions:
\begin{eqnarray}
    \alpha = 0~&~{\rm and}~&~\delta < \beta < 0~~{\rm and}~~\gamma=0~,\\
    \alpha = 0~&~{\rm and}~&~\delta - \gamma X_{\rm min}< \beta <
     0~~{\rm and}~~\gamma>0~,\\
    \alpha = 0~&~{\rm and}~&~\delta < \beta < - \gamma X_{\rm min}~~{\rm
     and}~~\gamma<0~. 
\end{eqnarray}
In any case, $t_{\pm}(x)$ have no first order term 
and the negative second order term with respect to $\ln(x/x_0)$.
Not only parameter $\beta$ but also $\delta$ is necessary for the solution.
We will apply the simulation to this analysis in the next work.

As other application of the simulation,
the detailed balance is expected to be extended to the detailed quasi-balance \cite{Ishikawa1}
observed in the assessed value of land in Japan \cite{Kaizoji, Web}.
If these applications are studied adequately, we can manage the
risk of these quantities by using knowledge of this method.




\section*{Acknowledgements}
We would like to thank the Yukawa Institute for Theoretical 
Physics at Kyoto University,
where this work was initiated during the YITP-W-07-16 on
``Econophysics III--Physical Approach to Social and Economic Phenomena--'',
and especially to 
Dr~Y. Soejima for the important comments about our work.
Thanks are also due to  Dr.~A. Sato
for a lot of useful discussions.
This work was supported in part
by a Grant--in--Aid for Scientific Research (C) (No.~20510147) 
from the Ministry
of Education, Culture, Sports, Science and Technology, Japan.

%

\newpage



\end{document}